\RequirePackage{fix-cm}
\documentclass[smallextended]{svjour3}       
\smartqed  
\usepackage{graphicx}
\begin{document}

\title{\textit{\textbf{Herschel}} SPIRE Fourier Transform Spectrometer: Calibration of its Bright-source Mode\thanks{{\it Herschel} is an ESA space 
observatory with science instruments provided by European-led Principal Investigator consortia and with important 
participation from NASA.}
}


\titlerunning{SPIRE FTS bright-source mode}        

\author{Nanyao Lu  \and
        Edward T. Polehampton \and 
        Bruce M. Swinyard \and
	Dominique Benielli  \and
	Trevor Fulton \and
	Rosalind Hopwood \and
	Peter Imhof \and
	Tanya Lim \and
	Nicola Marchili \and
	David A. Naylor \and
	Bernhard Schulz \and
	Sunil Sidher \and
	Ivan Valtchanov
}


\institute{
N. Lu \at
NHSC/IPAC, 100-22 Caltech, Pasadena, CA 91125, USA \\
\email{lu@ipac.caltech.edu}           
\and
E. T. Polehampton \at
RAL Space, Rutherford Appleton Laboratory, Didcot, Oxfordshire, OX11 0QX, UK, and \\
Institute for Space Imaging Science, Department of Physics \& Astronomy, University of Lethbridge, Lethbridge, AB T1K3M4, Canada
\and
B. M. Swinyard \at
RAL Space, Rutherford Appleton Laboratory, Didcot, Oxfordshire, OX11 0QX, UK, and \\
Dept. of Physics \& Astronomy, University College London, Gower St, London, WC1E 6BT, UK
\and
D. Benielli \at
Aix Marseille Universit\'e, CNRS, LAM (Laboratoire d'Astrophysique de Marseille) UMR 7326, 13388, Marseille, France
\and
T. Fulton \at
Institute for Space Imaging Science, Department of Physics \& Astronomy, University of Lethbridge, Lethbridge, AB T1K3M4, Canada
\and
R. Hopwood \at
Physics Department, Imperial College London, South Kensington Campus, SW7 2AZ, UK
\and
P. Imhof \at
Bluesky Spectroscopy Lethbridge University, Lethbridge, Canada
\and
T. Lim \at
RAL Space, Rutherford Appleton Laboratory, Didcot OX11 0QX, UK
\and
N. Marchili \at
Universit\'a di Padova, I-35131 Padova, Italy
\and
D. A. Naylor \at
Institute for Space Imaging Science, Department of Physics, and \\
Astronomy Department, University of Lethbridge, Lethbridge, AB, Canada, T1K 3M4
\and
B. Schulz \at
NHSC/IPAC, 100-22 Caltech, Pasadena, CA 91125, USA
\and
S. Sidher \at
RAL Space, Rutherford Appleton Laboratory, Didcot, Oxfordshire, OX11 0QX, UK
\and
I. Valtchanov \at
Herschel Science Centre, ESAC, P.O. Box 78, 28691 Villanueva de la Ca\~nada, Madrid, Spain
}


\maketitle

\begin{abstract}
The Fourier Transform Spectrometer (FTS) of the Spectral and Photometric Imaging
REceiver (SPIRE) on board the ESA \textit{Herschel} Space Observatory has
two detector setting modes: (a) a nominal mode, which is optimized for
observing moderately bright to faint astronomical targets, and (b) a bright-source 
mode recommended for sources significantly brighter than 500$\,$Jy, within 
the SPIRE FTS bandwidth of 446.7-1544$\,$GHz (or 194-671$\,$microns in wavelength), 
which employs a reduced detector responsivity and out-of-phase 
analog signal amplifier/demodulator.  We address in detail the calibration 
issues unique to the bright-source mode,  describe the integration of 
the bright-mode data processing into the existing pipeline for the nominal
mode, and show that the flux calibration accuracy of the bright-source mode
is generally within 2\% of that of the nominal mode, and that the bright-source 
mode is 3 to 4 times less sensitive than the nominal mode.

\end{abstract}

\section{Introduction}
\label{sec1}
The Spectral and Photometric Imaging REceiver (SPIRE; Griffin et al.~2010) is one of 
three focal-plane instruments on board the ESA {\it Herschel} Space Observatory 
({\it Herschel}; Pilbratt et al. 2010).  
It contains an imaging photometric camera and an imaging Fourier Transform 
Spectrometer (FTS).  The SPIRE FTS employs two detector
arrays of spider-web neutron transmutation doped (NTD) bolometers
(Bock et al. 1998), biased by
a sinusoidal AC voltage with a $160\,$Hz frequency:
a short-wavelength array (SSW) of 37 bolometers covering 959.3-1544 GHz
in frequency (194-313 $\mu$m in wavelength) and a long-wavelength array (SLW) 
of 19 bolometers covering 446.7-989.4 GHz (303-671 $\mu$m). 
Two detector-setting modes are available: (a) a nominal mode with the detector 
arrays optimally biased (by voltages of amplitude of 36 and 31 mV for SSW and 
SLW, respectively) to achieve the highest detection sensitivity, 
and (b) a bright-source mode  with the detectors subject to a much higher bias 
voltage (of amplitude of 176.4 mV for both SSW and SLW) to yield a 
reduced detector responsivity.  In the bright-source mode, the analog 
square-wave amplifier/demodulator, as sketched in Fig.~1\footnote{This figure was 
adapted from Schulz et al.~(2008), based on 
the document of {\it The SPIRE Analogue Signal Chain and Photometer Detector
Data Processing Pipeline}, available at 
http://herschel.esac.esa.int/twiki/pub/Public/SpireCalibrationWeb/Phot$_-$Pipeline$_-$Issue7.pdf},
is further tuned to be about 70 and 68 degrees, for SSW and SLW detectors, respectively,
out of phase with the detector signal to further reduce the chance of 
saturating the analog-to-digital converter.  As a result, the bright-source
mode results in a much larger dynamic range in flux, allowing for sources 
as bright as 25,000$\,$Jy to be observed without serious saturation.
In comparison, the nominal mode is recommended for sources fainter than 
${\sim}500\,$Jy within the SPIRE FTS bandwidth.

\begin{figure}
\centering
\includegraphics[width=0.80\textwidth]{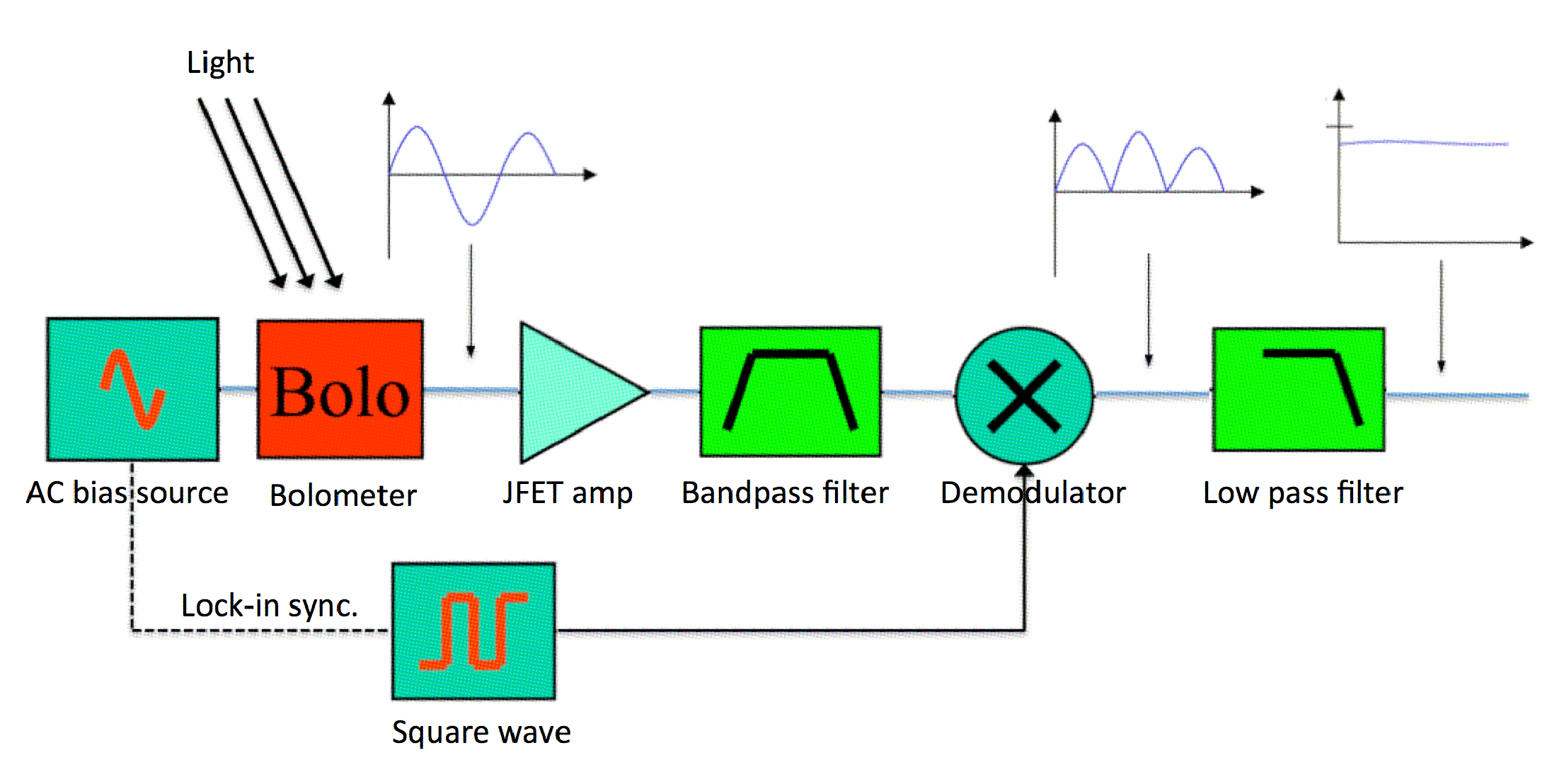}
\caption{An illustration of the first part of the SPIRE FTS signal chain:
The AC signal from the bolometer passes through the JFET amplifier, a bandpass filter 
that filters out the DC signal, a square-wave amplifier/demodulator that 
is normally locked in phase with the bolometer bias voltage to turn the negative 
part of the signal into positive, and finally a low pass filter
that generates a slowly varying DC signal.  In the bright-source mode, 
the square-wave demodulator is de-phased from the bolometer signal to further 
damp the output DC signal.
}
\label{fig1}    
\end{figure}

The overall strategy for the in-flight calibration of the SPIRE FTS is outlined 
in Swinyard et al.~(2010).  The actual calibration and pipeline implementation
of the FTS nominal mode are described in detail by Swinyard et al.~(2013) and 
Fulton et al.~(2013; 2010), respectively.  The bright-source mode calibration went
through a major upgrade in March 2013 [coinciding with Version 11 of {\it Herschel} 
Interactive Data Processing Environment (HIPE); Ott (2010)]. Prior to HIPE 11, the data of 
the bright-source mode was processed in the temperature domain, with some 
residual bolometer nonlinearity passed through to the end of the pipeline.
Starting in HIPE 11, we adopted a new full nonlinearity correction scheme and 
integrated the bright-source mode data processing directly into the existing 
pipeline for the nominal mode.  As a result, the agreement between the bright-source
and nominal mode flux calibrations has been improved from within ${\sim}10\%$ in HIPE 10
to within ${\sim}2\%$ in HIPE 11.   In this paper, 
we address those calibration issues unique to the bright-source mode (\S2),  
describe how the data obtained in the bright mode is folded into the nominal-mode 
pipeline for data processing (\S3), demonstrate that the bright-source mode 
flux calibration accuracy is within about 2\% of that of the nominal mode (\S4), 
and finally summarize our results (\S5).

\section{Calibration Scheme for the Bright-source Mode}
\label{sec2}

The calibration strategy for the bright-source mode is to utilize the calibration
products derived for the nominal mode wherever possible, in order to minimize 
the overall FTS calibration effort and keep the pipeline simple and robust.  
In this approach, the bright-source mode needs only the following unique 
calibration products or procedures: (1) a phase-related gain correction factor, 
$G_{\rm phase}$, for the out-of-phase 
analog amplifier/demodulator, (2) a detector nonlinearity correction calibration 
product, (3) a zero-point, DC-type gain correction factor, $G_0$, which aligns 
the linearized signal scale of the bright-source mode to that of the nominal mode, 
and (4) a possible frequency-dependent gain factor, $G_f$.  The last one may result 
from effects such as a dependence of bolometer response time constant on 
the bias voltage.  We address each of these issues in more detail below.

\subsection{Phase-related Gain Correction}
\label{sec2.1}

The square-wave analog amplifier/demodulator is locked in phase with 
the bolometer AC signal in the nominal mode, but is intentionally kept
at $\phi_{\rm diff}$ out of phase in the bright source mode; where
$\phi_{\rm diff}$, measured for each detector, varies slightly for 
different detectors of the same bolometer array, and is around 70 and 
68 degrees for SSW and SLW detectors, respectively. 
As discussed in Swinyard et al.~(2013), the effective R-C circuit of
the bolometer JFET amplifier and harness introduces a gain factor, $G_{\rm cab}$, 
and a signal phase shift, $\phi_t$, as follows:
\begin{equation}
G_{\rm cab} = \sqrt{{1 \over 1 + \omega^2_{\rm cr}}}, 
\end{equation}
\begin{equation}
\phi_t = {\rm atan}(\omega_{\rm cr}) + \phi_{\rm off},
\end{equation}
where $\phi_{\rm off}$ is a constant phase offset ($= 11.4$ and $13.6$ degrees for SSW
and SLW, respectively) and $\omega_{\rm cr} = 2\pi\omega_{\rm bias}R_{\rm tot}C_{\rm H}$,
with $\omega_{\rm bias}$ being the detector bias frequency, $R_{\rm tot}$ is
the total resistance of the bolometer readout circuit, including 
the bolometer itself and the load resistors, and $C_{\rm H}$ ($= 20\,$pF) 
is the cable capacitance of the FTS readout system.  Since the bolometer 
resistance depends on the optical load, so does $G_{\rm cab}$.  In practice, 
$G_{\rm cab}$ and $\omega_{\rm cr}$ are determined in an iterative way.
The gain factor related to the adjusted phase is given by:
\begin{equation}
G_{\rm phase} = \cos(\phi_{\rm diff} - \phi_t).
\end{equation}
This $G_{\rm phase}$ is divided into each voltage sample for the bright-source 
mode at the engineering data conversion stage in the pipeline. Typically,
$G_{\rm cab}$ is always close to unity and $G_{\rm phase}$ is on the order of 0.5 
(e.g., it is around 0.54 for SSWD4 and around 0.67 for SLWC3).

\subsection{Detector Nonlinearity Correction}
\label{sec2.2}

Since the nonlinear responsivity of a bolometer depends on its bias voltage,
it is necessary to derive a separate nonlinearity correction calibration
product for each of the two detector-setting modes.  
For the SPIRE bolometers, the signal linearization can be done 
in an analytic way as, following Swinyard et al.~(2013) and Bendo 
et al.~(2013),
\begin{equation}
V' = K_1(V_m - V_0) + K_2\,\ln({V_m - K_3 \over V_0 - K_3}),
\end{equation}
where $V_m$ and $V'$ are the observed and linearized bolometer voltages, 
respectively, $V_0$ is an {\it arbitrary} reference voltage, and $K_1$, 
$K_2$ and $K_3$ are the parameters characterizing the detector nonlinearity.
For the nominal mode, these K parameters were derived from a physical 
bolometer model (e.g., Sudiwala et al. 2002; Woodcraft et al. 2002) using
a bolometer analysis package developed at the NASA {\it Herschel} Science Center
(Schulz et al.~2005) with both laboratory and in-flight 
measured detector parameters (Nguyen et al.~2004). 
For the bright-source 
mode, which requires nonlinearity correction over a much larger flux 
range, these K parameters were determined directly from the calibration
data taken on flashes of the SPIRE internal photometric calibrator (PCAL;  
Pisano et al.~2005).   Each astronomical FTS observation contains 9 
pairs of PCAL flashes on top of the background emission at the position of 
the target. Some dedicated PCAL calibration observations were also obtained
in order to expand the background flux coverage.   

For each PCAL flash of power off and power on, we can write
\begin{equation}
1/\delta V_m  = K_1 + K_2/(V^{\rm off}_m - K_3),
\end{equation}
where $\delta V_m$ is the instantaneous bolometer voltage change when 
the PCAL power is turned on and $V^{\rm off}_m$ is the voltage reading just
before the PCAL power is turned on.  We can write eq.~(5) because
the PCAL power and illumination pattern remain fixed over the entire 
mission (as well as between the nominal and bright-source modes) and
because an arbitrary common scaling factor is allowed for $K_1$ and 
$K_2$. (This scaling factor gets folded into the zero-point gain correction 
factor in \S2.3.)

As examples, Fig.~2 shows two independent sets of PCAL flashes on 
the detector SSWD4.  The left-hand side plot represents a 
set of PCAL flashes typically 
seen in an astronomical observation.  Note that there is a 
slight downward signal drift when the PCAL power is on, illustrating 
a possible heat input to the detectors from the PCAL power. 
One of our dedicated PCAL calibration 
observations towards the Galactic center is shown in the right-hand plot
to illustrate a typical PCAL
observation when the telescope was pointed at a bright discrete source.  
The strong baseline drift over the on-off cycle was a result of the jitter
in the telescope pointing.   Our PCAL data reduction pipeline 
module fit a linear function independently to each on and off signal 
plateau (after excluding a certain percentage of the data points at 
the beginning and end of each plateau; see the SPIRE pipeline description 
document\footnote{The SPIRE pipeline description 
document, available at http://herschel.esac.esa.int/twiki/bin/view/Public/SpireCalibrationWeb,
will be updated to reflect the PCAL data reduction algorithm described here.}
for more details) and determines from these fits $V^{\rm off}_m$ 
and $V^{\rm on}_m$ ($= V^{\rm off}_m + \delta V_m$) per PCAL flash pair.
Finally, for each PCAL observation, the median $V^{\rm off}_m$ and 
$\delta V_m$ values over all of its PCAL flash pairs were derived for 
use in our detector nonlinearity characterization.

\begin{figure}
\includegraphics[width=0.48\textwidth]{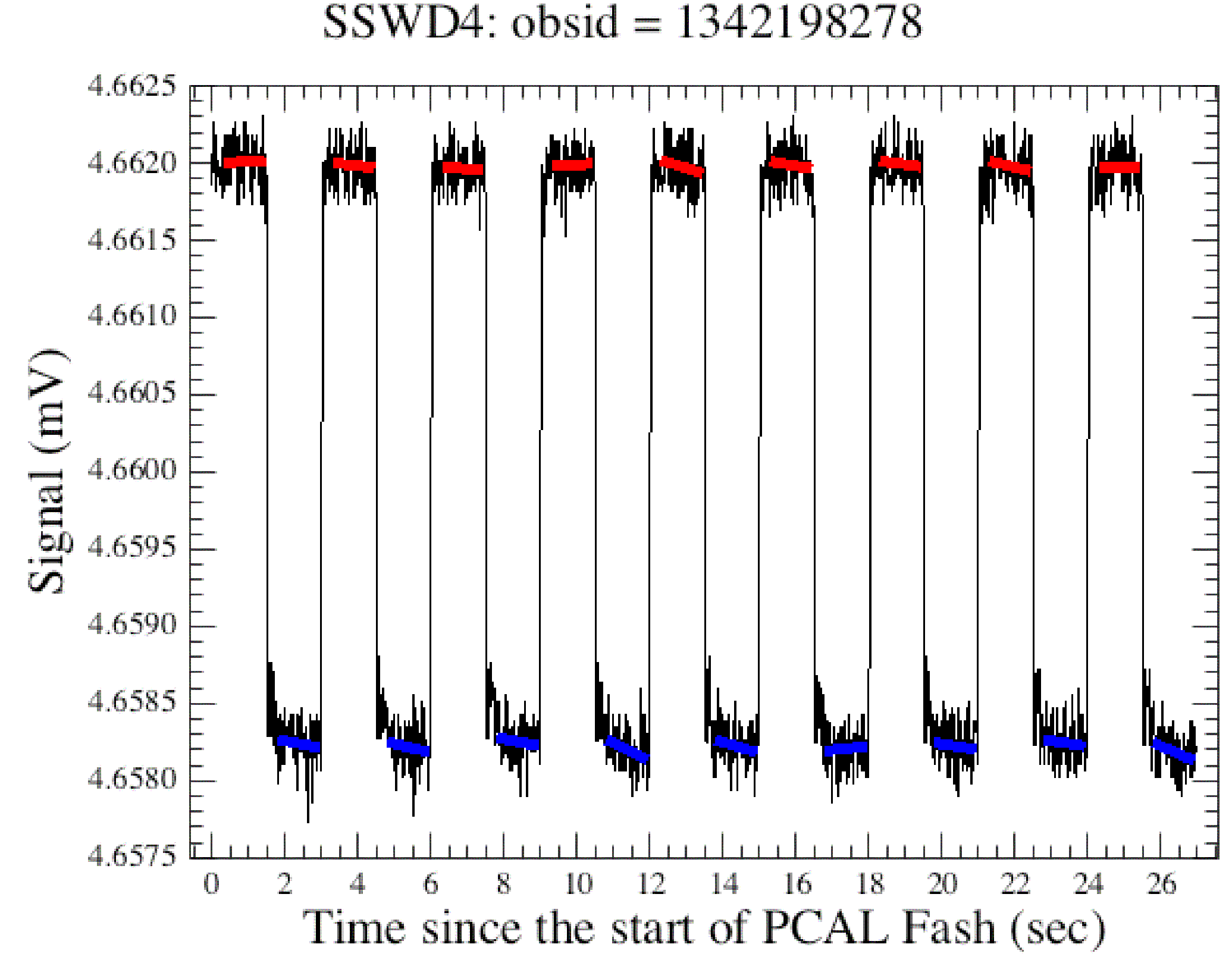}
\hspace{0.01\textwidth}
\includegraphics[width=0.48\textwidth]{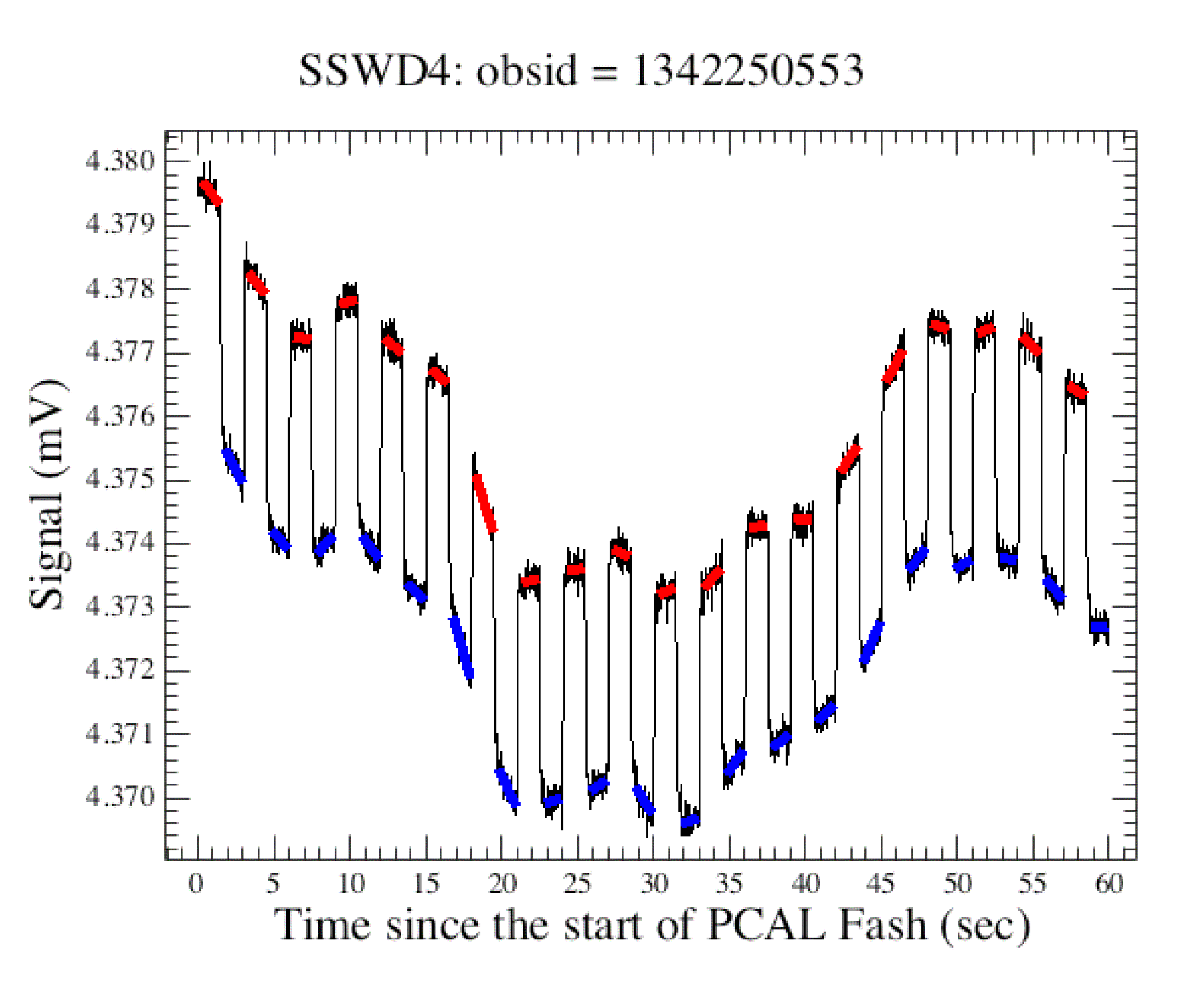}
\caption{Two examples of PCAL flashes on the detector SSWD4. 
The actual data samples are shown in black curves. The fits to the power-off 
and power-on signal 
plateaus are shown as red or blue lines, respectively.
}
\label{fig2}    
\end{figure}

Fig.~3 illustrates our fitting of eq.~(5) to the bright-source mode PCAL data 
pairs (i.e., $V^{\rm off}_m$ and $\delta V_m$) we accumulated over the entire mission
for the two central detectors, SSWD4 and SLWC3.  The voltage coverage along
the X-axis ranges from dark sky observations (at the high voltage end) to 
those from two Saturn observations. It is evident that the data points are still 
sparse at the low voltage end, leading to possibly a lower flux calibration accuracy
for bright targets such as Saturn.  For each detector, we defined a voltage range
of $V^{\rm off}_m{\rm (min)}$ to $V^{\rm off}_m{\rm (max)}$, within which 
the nonlinearity correction based on the fit is deemed to be valid.  The value of
$V^{\rm off}_m{\rm (min)}$ is set to 5\% below the smallest voltage sample observed 
and that of $V^{\rm off}_m{\rm (max)}$ to 3\% above the largest voltage sample
we have.  If we compare the PCAL $\delta V$ values on the same background source 
between the bright-source and nominal modes (for the nominal mode counterpart to
Fig.~3 here, see their Fig.~5 in Swinyard et al.~2013), in general, the 
detector responsivity in the bright-source mode is about a quarter of that in 
the nominal mode.

\begin{figure}
\includegraphics[width=0.48\textwidth]{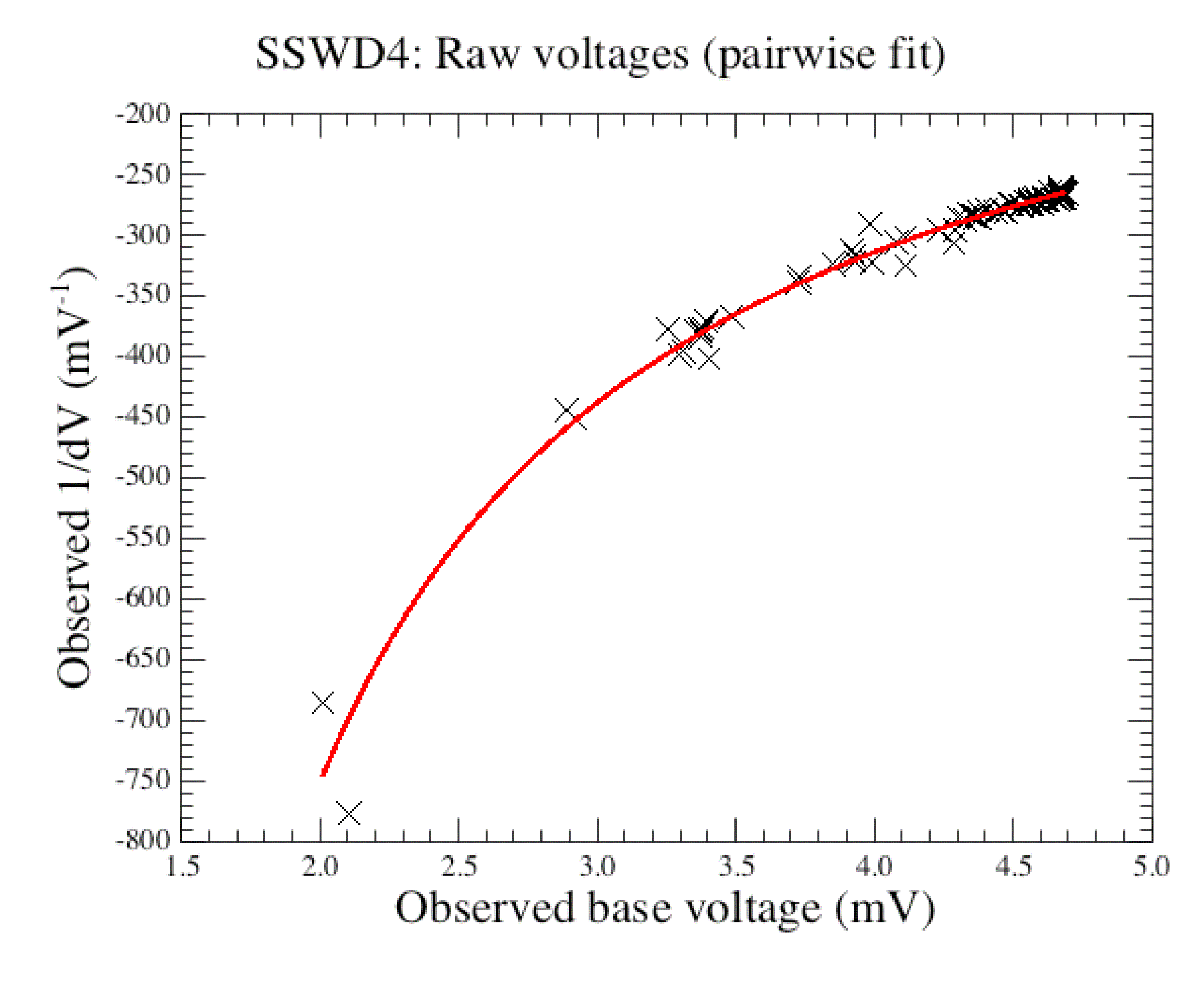}
\hspace{0.01\textwidth}
\includegraphics[width=0.48\textwidth]{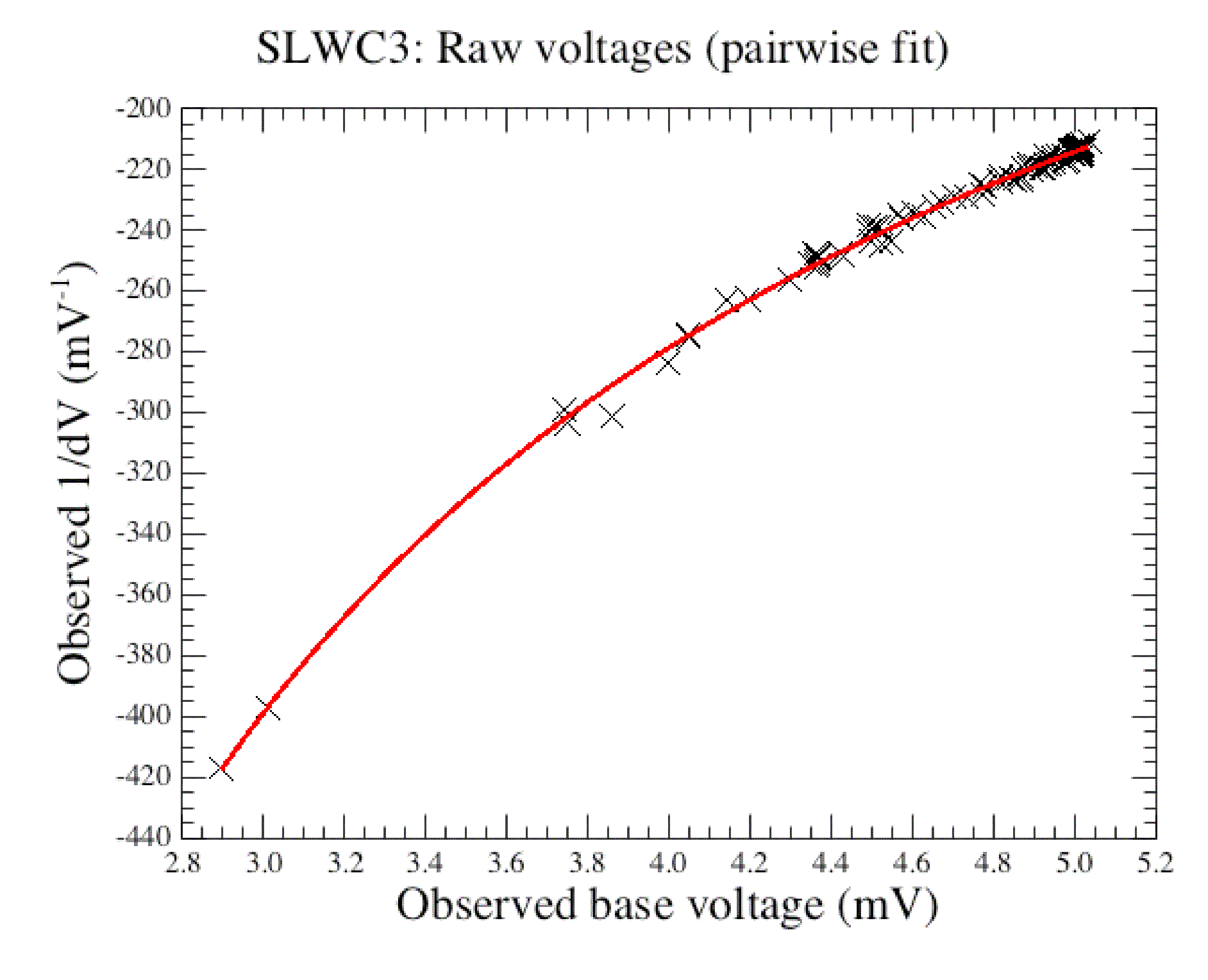}
\caption{Bright-source mode bolometer nonlinear responsivity fits of eq.~(5) to 
the PCAL data for the two central
detectors, SSWD4 (on the left-hand side) and SLWC3 (on the right).  
The median results from individual sets 
of PCAL flashes are shown in black crosses.  The two data points with the lowest 
observed base voltages are from PCAL flashes on Saturn.  The best fit is shown 
as a red curve.}
\label{fig3}      
\end{figure}

\subsection{Zero-point Gain Correction}
\label{sec2.3}

As an example, Fig.~4 shows the linearized PCAL voltages [via eq.~(4)] for 
the detector SSWD4 in the nominal mode (on the left-hand side) and 
bright-source mode (on the right).  These plots also illustrate that, for 
the majority of the detectors,  the typical sample standard deviation for 
the linearized PCAL signals is of the order of $2\%$ for the bright-source 
mode and is less than $1\%$ for the nominal mode.  
Since the linearized voltage is proportional
to the optical load on the detector, and the PCAL power and illumination 
pattern was kept the same for both detector modes, the linearized 
voltage ratio of the nominal mode to the bright-source mode gives a 
zero-point gain scaling factor, $G_0$, from the bright-source to nominal 
mode.  The resulting $G_0$ varies between 4.1 and 5.4, depending on 
specific detector.

\begin{figure}
\includegraphics[width=0.48\textwidth]{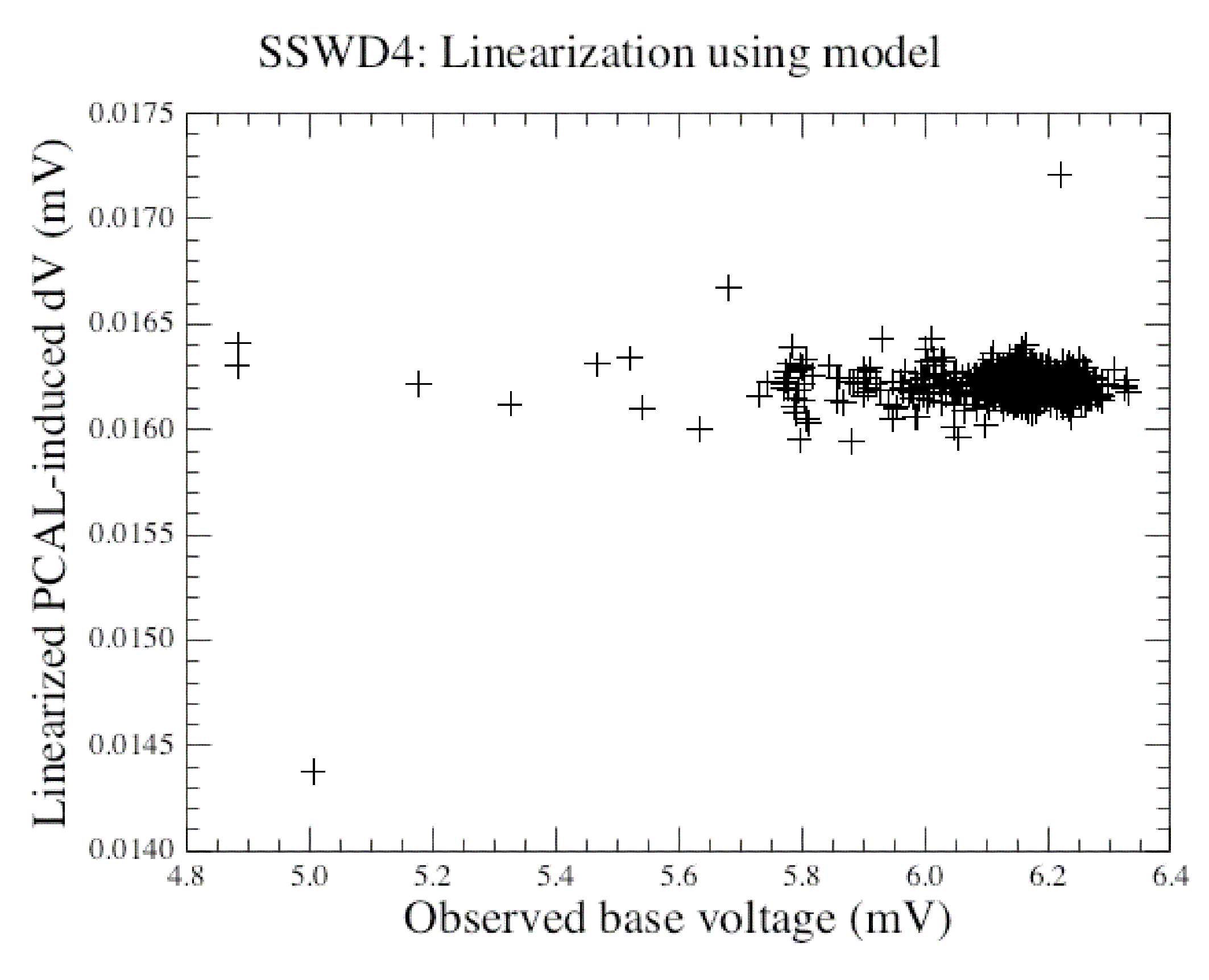}
\hspace{0.01\textwidth}
\includegraphics[width=0.48\textwidth]{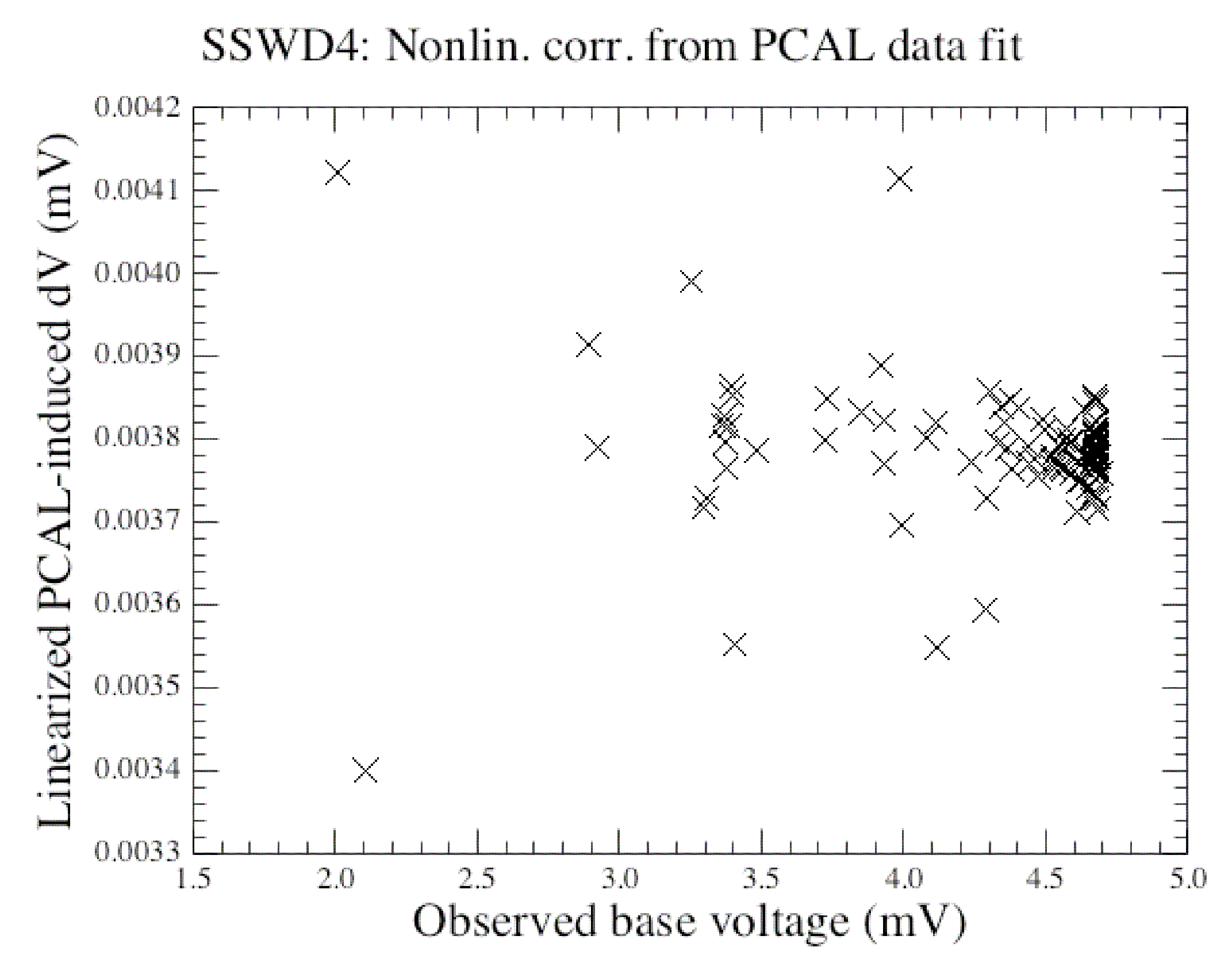}
\caption{Linearized PCAL signals for the central detector SSWD4, from 
the nominal model on the left-hand side and from the bright-source mode 
on the right.}
\label{fig4}     
\end{figure}

\subsection{Frequency-dependent Gain Correction}
\label{sec2.4}

In addition to $G_0$, which was derived from low-frequency PCAL signal time lines, 
we also expect an additional frequency-dependent scaling factor between the two
detector setting modes to account for the higher frequency signal modulations in 
interferograms.  This can arise from the fact that bolometer time constant 
depends on the bias voltage used.   
While there is a correction for the finite bolometer time constant implemented 
in the pipeline, any residual effect from imperfect correction could lead to some 
spectral shape distortion.  In the nominal mode, this potential residual spectral
shape distortion is simply corrected for at the flux calibration step in the pipeline.  
To make use of the same flux calibration product for the bright-source mode, we 
introduced a frequency-dependent gain factor, $G_f$, which is to be applied to 
bright-source mode spectra.

Fig.~5 shows a number of pair-wise ratios of the nominal to bright-source mode 
for the two central detectors, SSWD4 and SLWC3, using dark sky spectra taken 
in the low spectral resolution configuration.  
Only the zero-point gain correction, $G_0$, has been applied to 
the bright-source data here.   Each pair of observations
were carried out close in time so that the telescope emission, which dominates 
the signal observed, remained unchanged over the observational pair.   Apart from 
an increased uncertainty at the low frequency end, where the removal of the instrument
emission, which is significant only near that end of SLW, introduces additional 
flux uncertainties, the ratios show approximately a linear dependence on frequency
for each detector array.  A linear fit was applied to the data of SSWD4 or SLWC3, 
resulting $G_f$ as a function of frequency.  Note that the correction associated
with $G_f$ is less than ${\sim}5\%$.  Similar fits were obtained for all the other
detectors.

\begin{figure}
\centering
\includegraphics[width=0.80\textwidth]{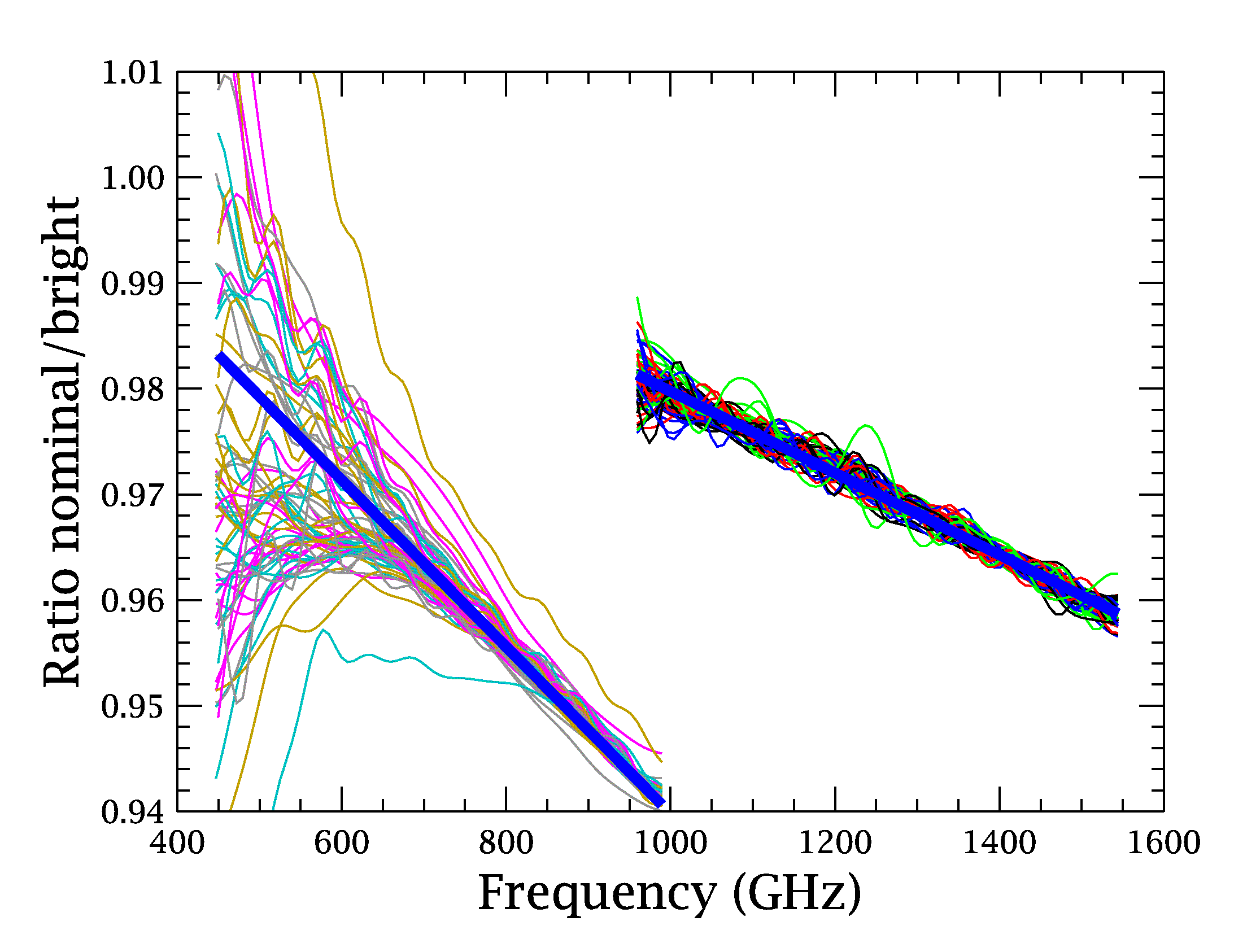}
\caption{Frequency dependency of pairwise dark spectrum ratios of the nominal 
mode to the bright-source mode for the two central detectors, SSWD4 and SLWC3. 
Only the zero-point gain correction, $G_0$, was
applied to the data of the bright mode here.  Independent data pairs 
are coded in different colors.  The spectra have been slightly smoothed in 
frequency to reduce noise.  The two thick blue lines are the best linear fits
to the data over SSWD4 and SLWC3, respectively.
}
\label{fig5}     
\end{figure}

\section{Pipeline Implementation}
\label{sec3}

Fig.~6 is a flow chart illustration on how the bright-source mode data processing
is folded into the standard nominal-mode pipeline. The bright-source 
mode pipeline processing is the same for the nominal mode except 
for the following three stages in the pipeline:  (a) The phase-related
gain correction 
option is turned on for the bright-source mode at the step of the engineering data 
conversion.  (b) While both detector modes share the sample nonlinearity 
correction module, they use separate nonlinearity calibration products.  
(c) After the Fourier transform,  there is an extra step for the bright-source mode, 
i.e., each spectrum is multiplied by a combined gain correction factor, 
$G_0\,G_f$,  which is a linear function of frequency for each detector.

\begin{figure}
\centering
\includegraphics[width=0.80\textwidth]{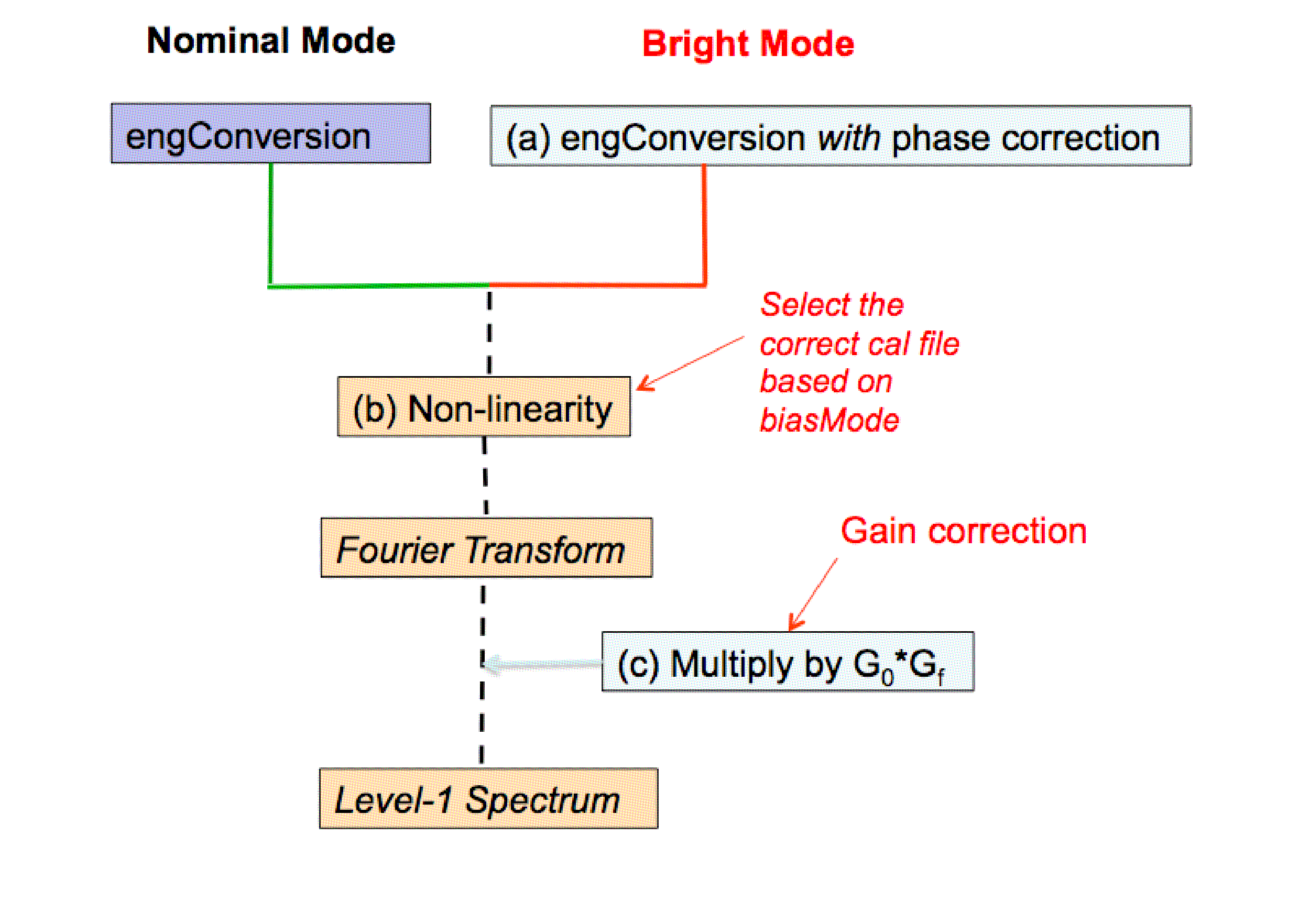}
\caption{A flow chart illustration on the integration of the bright-source mode 
data processing into the existing nominal mode pipeline.  The three processing 
stages where the two detector modes are treated differently are marked by (a), 
(b) and (c), respectively, and are described in the text.
}
\label{fig6}     
\end{figure}

\section{Calibration Results}
\label{sec4}

The validity and consistency of the calibration scheme described above can 
be studied by comparing the bright-source pipeline results with those from 
the nominal mode for some bright sources that are observable in both 
observing modes or with independent flux models of very bright celestial 
standards.
Fig.~7 checks the pipeline results of all the high-resolution dark sky 
observations taken in the bright-source mode over the entire mission.  
The two central detectors, SSWD4 and SLWC3, are shown here.  Individual 
spectra have been smoothed to reduce effects of noise and fringes. 
A dark observation is dominated by the warm telescope emission, with an
in-beam flux density of ${\sim}200-800\,$Jy over the whole FTS bandwidth.
A perfect flux calibration would 
yield a flat spectrum at 0 Jy for these observations as the telescope 
emission is removed in the pipeline.   It is evident that the mean from 
these dark sky 
spectra is close to 0 Jy.  The sample standard deviation is of the order 
of 0.5$\,$Jy, except for the low frequency end of SLWC3 where the scatter is 
somewhat elevated mainly due to the fact that the removal of the instrument
emission, which is significant only near that end of SLW, introduces
additional flux uncertainties.

\begin{figure}
\centering
\includegraphics[width=0.80\textwidth]{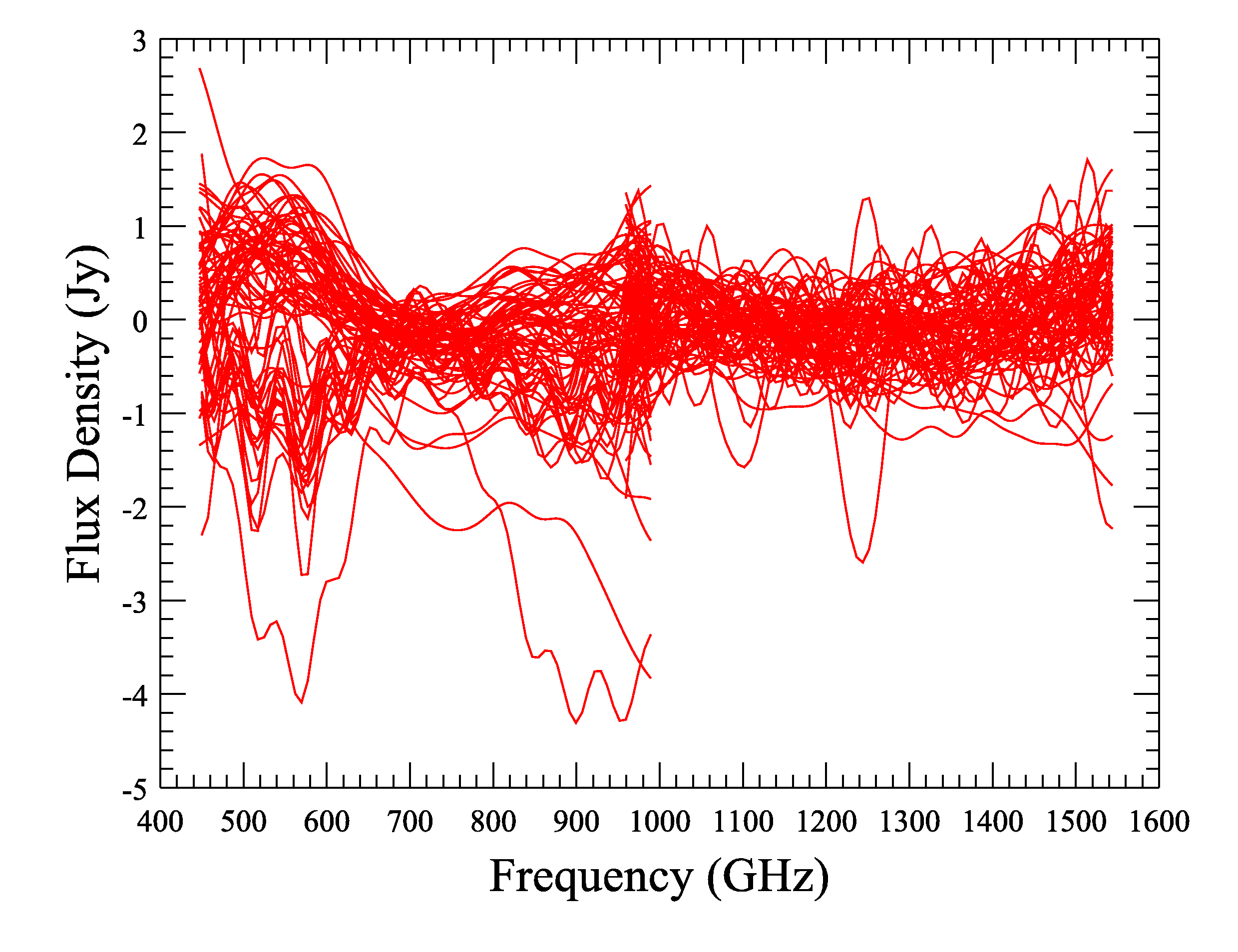}
\caption{Pipeline spectra of all the high-resolution dark sky observations 
made in the bright-source mode over the entire mission. Shown here are 
the two central detectors, SSWD4 and SLWC3.   Each spectrum has
been smoothed to reduce noise and fringes.
}
\label{fig7}     
\end{figure}

Fig.~8 shows the (smoothed) spectral ratios of the nominal mode to the bright-source mode 
for a few independent observations of Neptune and Uranus in the central detectors,
SSWD4 and SLWC3. 
Neptune and Uranus are the main photometric flux standards 
for SPIRE and span a flux-density range from a few tens of Jy to 220 and 500$\,$Jy, 
respectively,  within the SPIRE FTS bandwidth. It is evident that these spectral ratios 
are all within 2\%.

\begin{figure}
\includegraphics[width=0.48\textwidth]{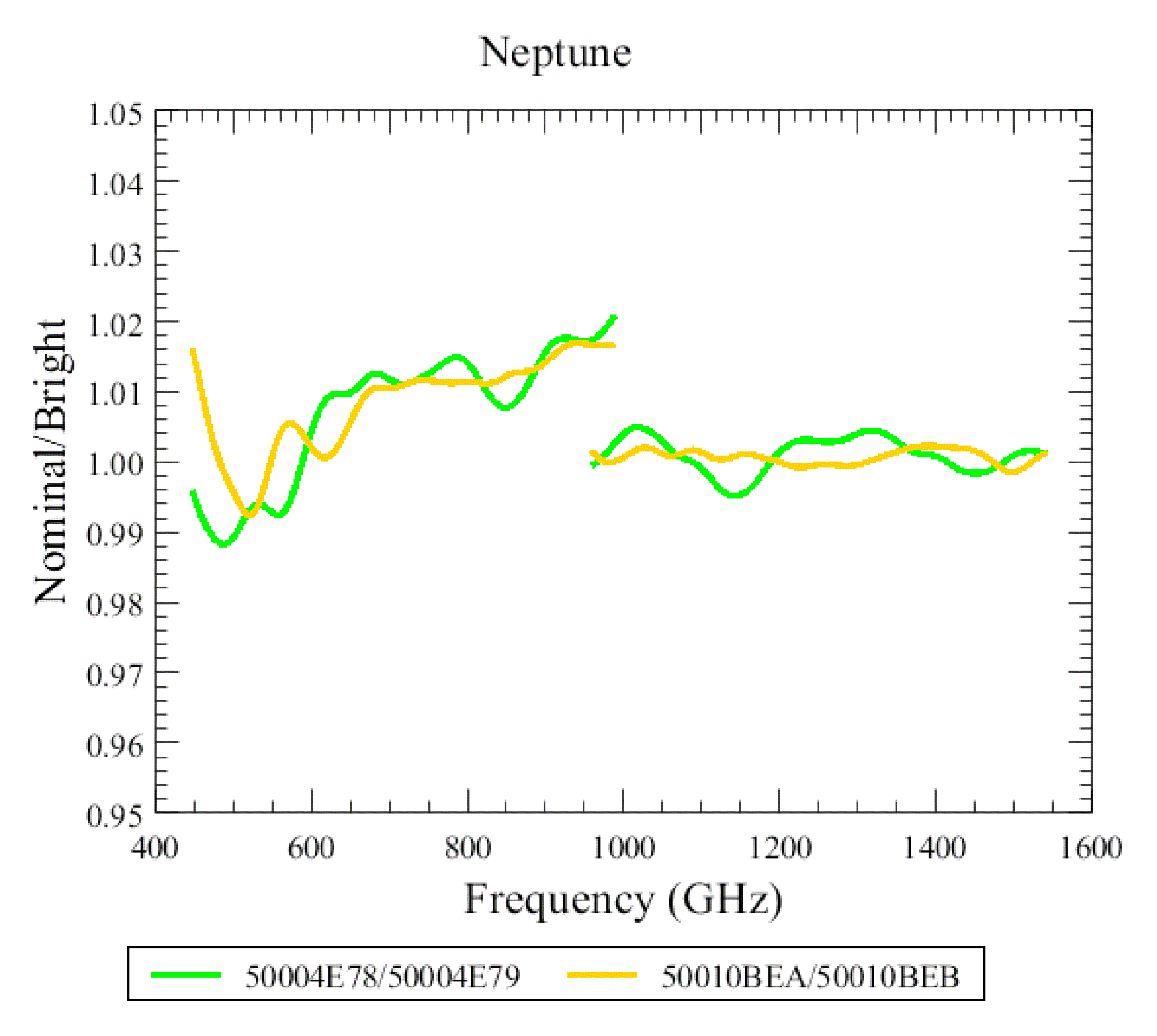}
\hspace{0.01\textwidth}
\includegraphics[width=0.48\textwidth]{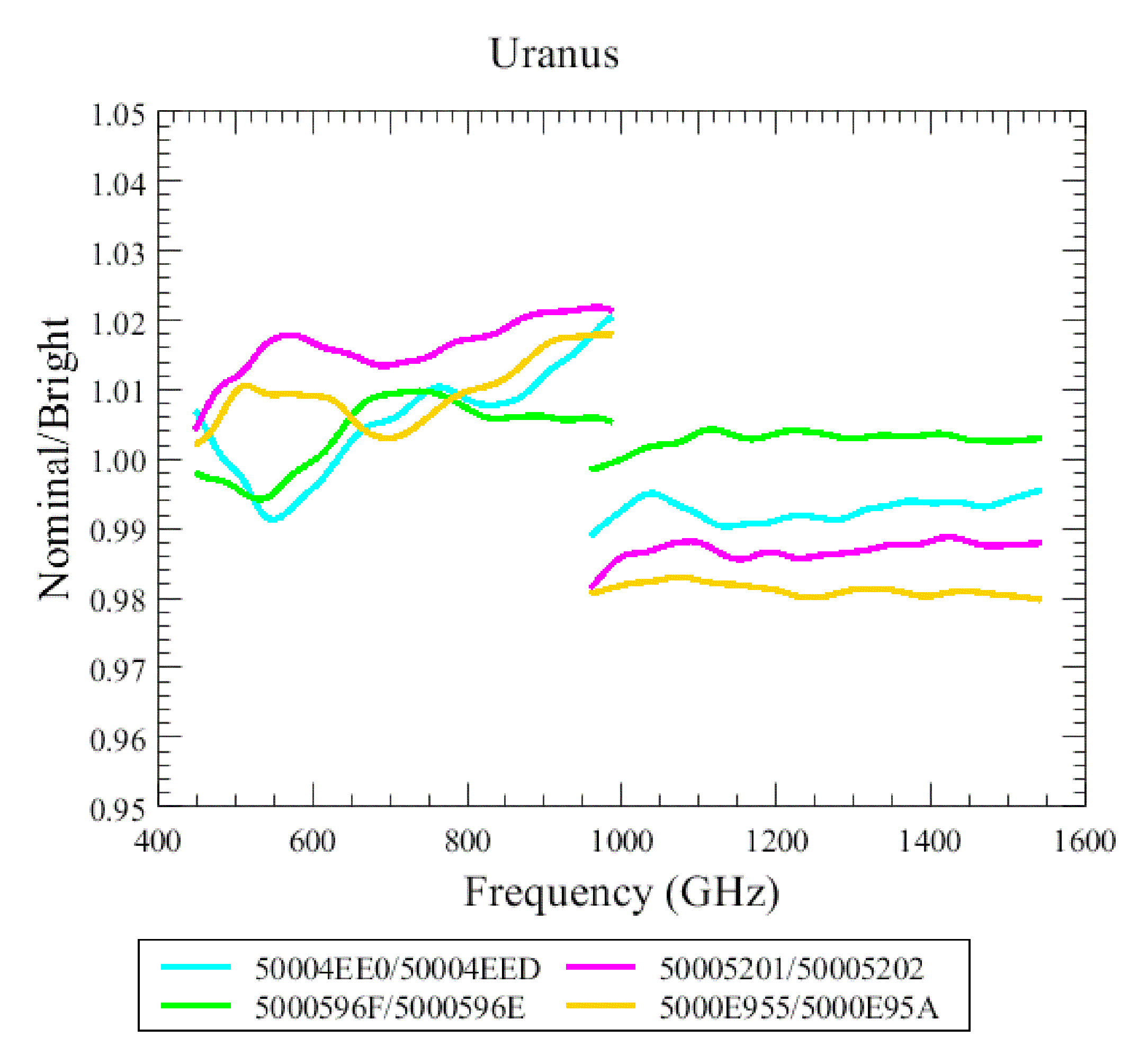}
\caption{Spectral ratios of the nominal mode to the bright-source mode for a few independent 
observations of Neptune (on the left-hand side) and Uranus (on the right) in the central 
detectors, SSWD4 and SLWC3.  Individual 
observations are coded in different colors.  All
the spectra have been smoothed to reduce noise.
}
\label{fig8}     
\end{figure}

Fig.~9 shows the (smoothed) spectral ratios of the nominal mode to the bright mode 
for two massive stars, Eta Car and AFGL$\,$2591.  Within the SPIRE
FTS beams and bandwidth, these two sources span a flux density range from a few 
tens of Jy to about 600 and 1,000$\,$Jy, respectively.  The flux differences 
between the bright-source and nominal modes are again within about 2\% here.

\begin{figure}
\centering
\includegraphics[width=0.80\textwidth]{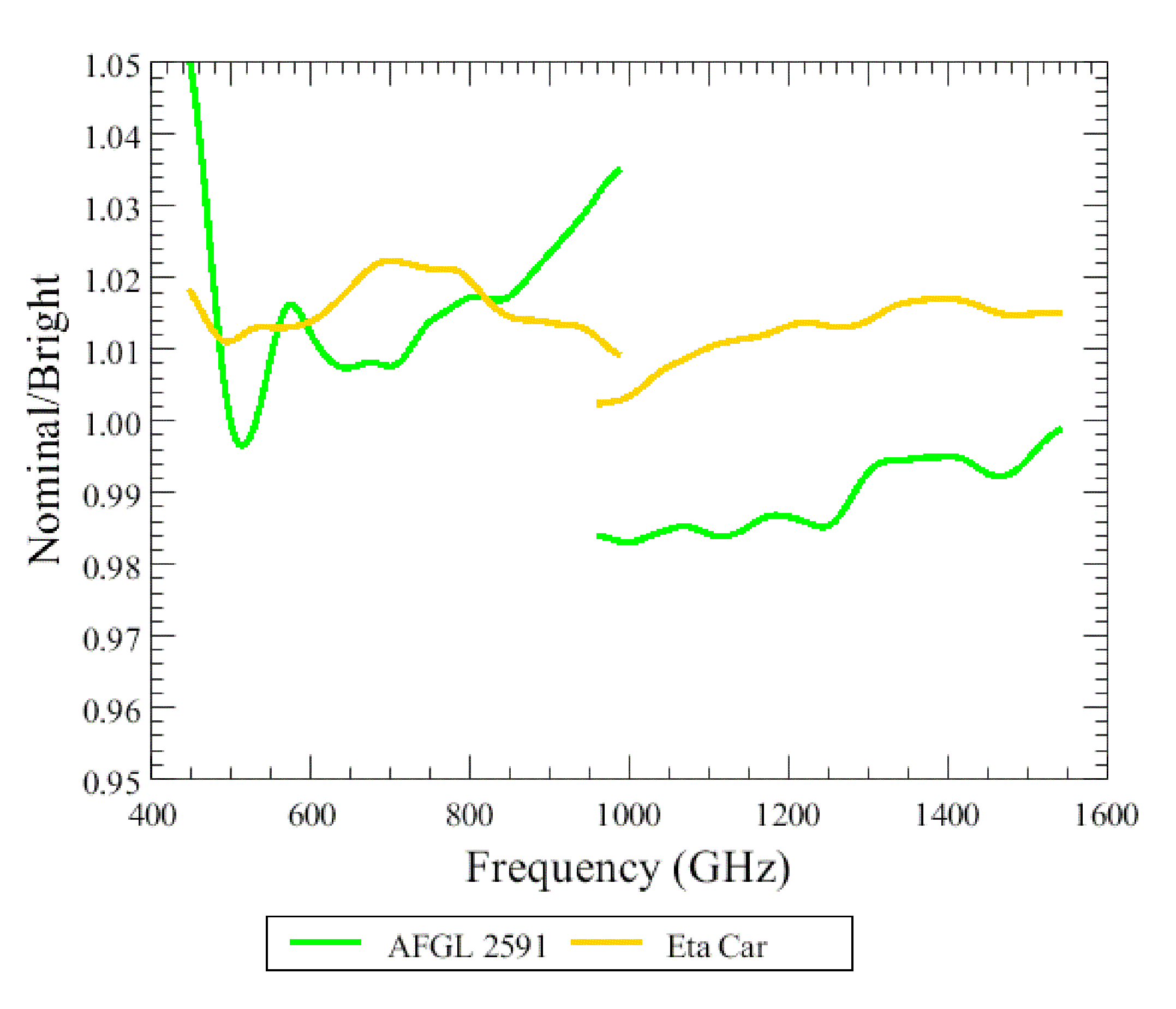}
\caption{Spectral ratios of the nominal mode to the bright mode for two massive 
stars, Eta Car (in green color) and AFGL$\,$2591 (in brown), observed in the central
detectors SSWD4 and SLWC3.   The spectra have been smoothed to reduce noise.}
\label{fig9}     
\end{figure}

For sources even brighter than those in Fig.~9, there is no longer any nominal 
mode data for comparison as severe saturation would have occurred.  For a few bright 
planets, there are reasonably accurate model spectra
available.  We compared our bright-source mode observations with the model spectra
for Mars and Saturn.   Fig.~10 compares the spectra from three independent, 
bright-source mode observations of Mars with the model-calculated flux densities 
at a few selected frequencies.   The latter data were derived from an online Mars brightness model 
provided by E.~Lellouch and H.~Amri, available at http://www.lesia.obspm.fr/perso/emmanuel-lellouch/mars/.
For the three observations, Mars was at different distances, hence had different 
apparent diameters (given in the figure caption) and fluxes.
The SPIRE spectra have been corrected for a finite angular size appropriate at
the time of the observation, using the SPIRE semi-extended source correction tool 
(Wu et al.~2013).  The agreement with the model fluxes is generally good to 
within a few percent.

The recommended flux density limit for the nominal mode is ${\sim}500\,$Jy, or roughly 
${\sim}$1,000$\,$Jy including the telescope background.   The bright-source mode 
suppresses voltage signals by a factor of ${\sim}8$ (with a factor of ${\sim}2$ from the de-phased amplifier 
and a factor of ${\sim}4$ from the reduced detector responsivity).  The corresponding 
targeted upper flux density limit for the bright-source mode is therefore on 
the order of 7,500$\,$Jy ($= 8 \times$1,000$\,$Jy less $500\,$Jy of the telescope 
background).  This is comparable to the flux density of Mars in Fig.~10 and 
encompasses the range seen in the vast majority of the bright-source mode science 
observations.  Only a few objects brighter than Mars, such as Saturn, were ever 
observed during the entire mission.

\begin{figure}
\centering
\includegraphics[width=0.80\textwidth]{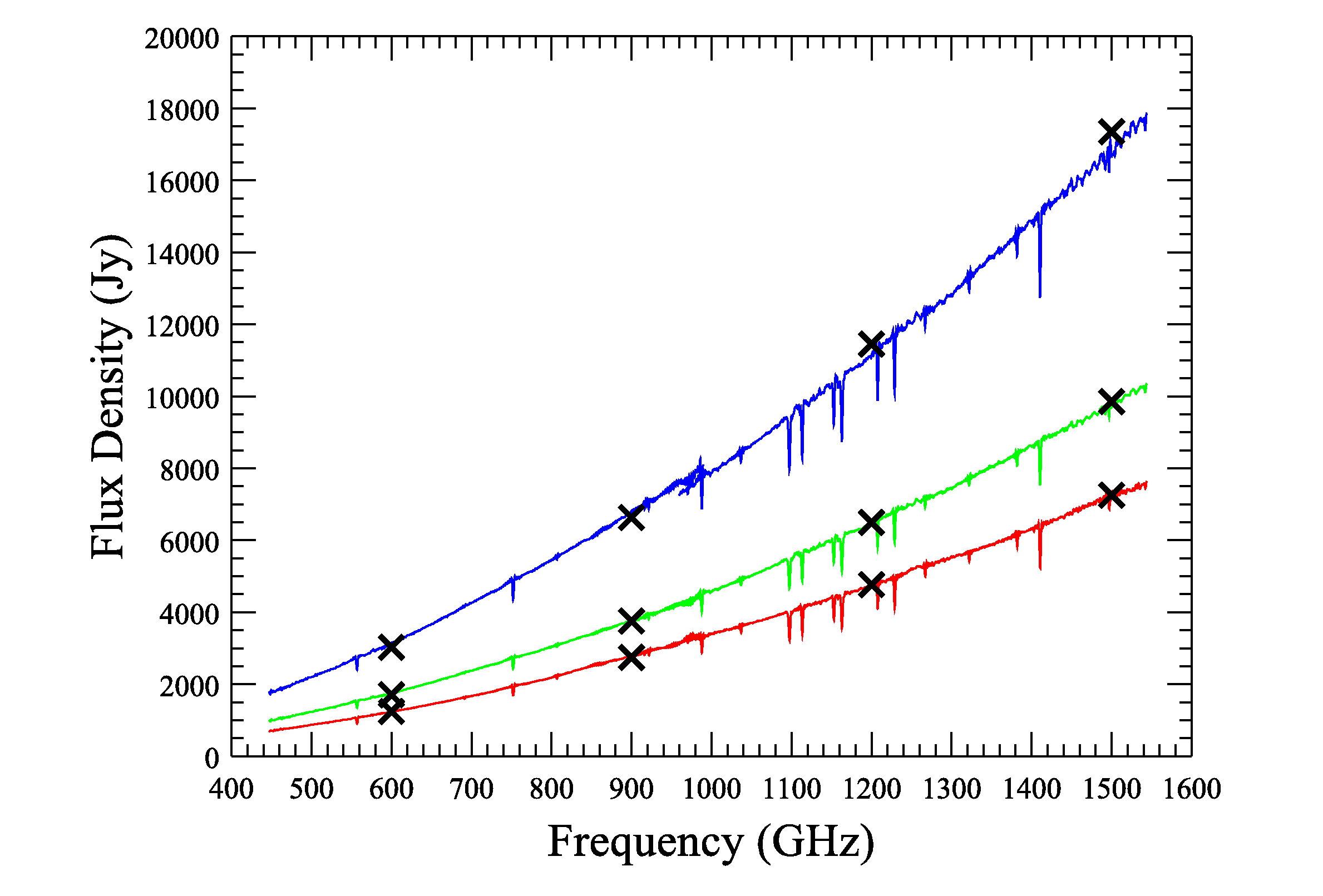}
\caption{Comparison of the spectra from three independent bright-mode observations 
of Mars with the model predicted flux densities (crosses) at a few selected frequencies. 
The three observations are identified by the following obs.~ID: 0x5000D22D
(or 1342231085; red curve, the angular size used is $\theta = 5.5''$), 0x50011293 
(or 1342247571; green curve, $\theta = 6.6''$), and 0x50010BD8 (or 1342245848; 
blue curve, $\theta = 8.8''$). Each SPIRE spectrum 
has been corrected for the finite angular size, $\theta$, of Mars appropriate at 
the time of the observation using the SPIRE semi-extended source correction tool.
}
\label{fig10}     
\end{figure}

Fig.~11 compares an observed SPIRE FTS spectrum in the bright-source mode with 
a model spectrum for Saturn.  The model spectrum is from Fletcher et al.~(2012)
after scaling their spectrum to the angular size ($= 17.4''$) of Saturn at 
the time of our bright-source mode observation.   Saturn is so bright that its 
induced bolometer voltages are at the low voltage end of the calibrated range 
for the detector nonlinearity correction, and its flux levels are significantly above 
the targeted upper flux density limit for the bright-source mode.   The flux calibration 
in this case is likely less accurate than in those of typical bright-source mode 
observations (i.e., fluxes up to that of Mars).  Nevertheless, the largest flux 
discrepancy in Fig.~11 is still less than 10\% at such extreme flux levels.  
A few other independent Saturn observations also confirm a flux uncertainty
of less than 10\%.

\begin{figure}
\centering
\includegraphics[width=0.80\textwidth]{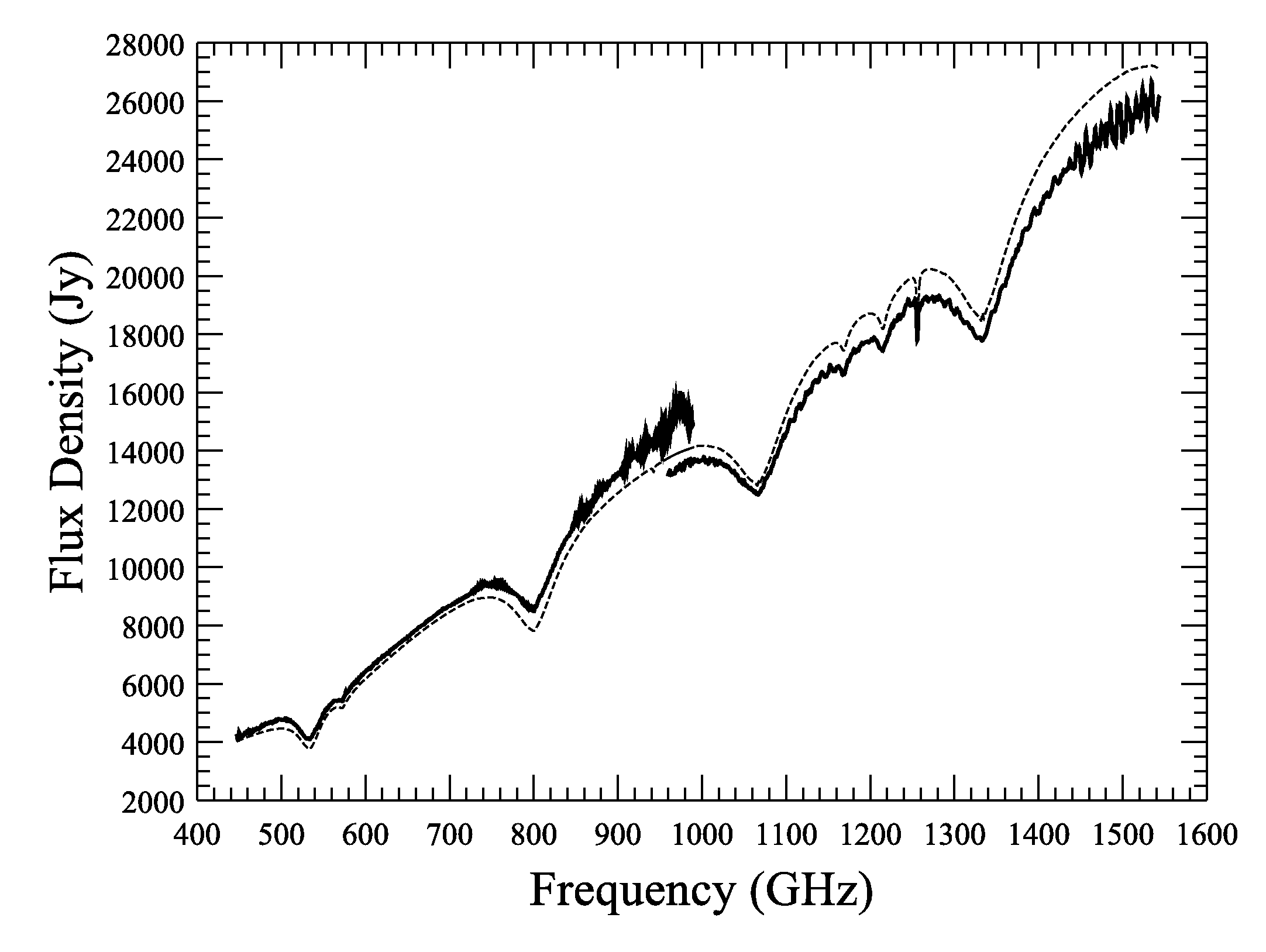}
\caption{Comparison of an observed spectrum (obs.~ID = 0x50011346 or 1342247750; 
thick curve) of Saturn with the model predicted spectrum (thin, dashed curve).  The SPIRE 
spectrum has been corrected for a 
finite angular size of $17.4''$ for Saturn at the time of the observation, using the SPIRE 
semi-extended source correction tool. 
}
\label{fig11}     
\end{figure}

Finally, Fig.~12 shows the average sensitivity ratio of the bright-source mode to the nominal
mode, based on the dark observations taken post Operational Day 1011.  The sensitivity prior to 
that day is very similar.  This sensitivity was derived from the observed spectral noise
of the dark observations.  For each dark spectrum, the 1-$\sigma$ r.m.s.~noise was calculated 
within each and every spectral bin of 50$\,$GHz.  The noise of a specific spectrum was 
further normalized to a reference integration time before the average sensitivity was finally
calculated for each observing mode.  The results show that the bright-source mode is about 
3 to 4 times less sensitive than the nominal mode.

\begin{figure}
\centering
\includegraphics[width=0.80\textwidth]{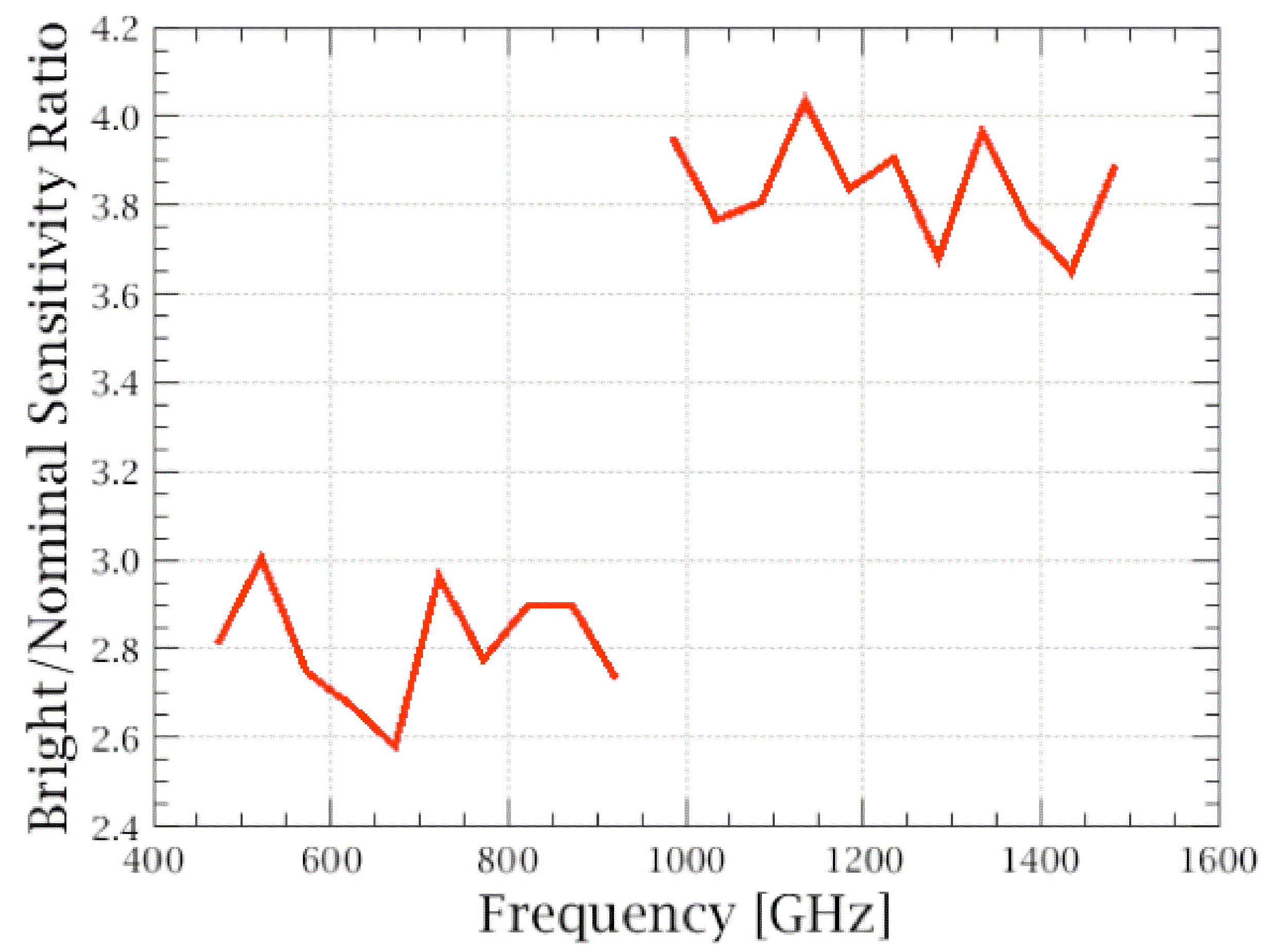}
\caption{Average sensitivity ratio of the bright-source mode to the nominal mode, based 
on all the dark observations post Operational Day 1011. It shows that the bright-source 
mode is about 3 to 4 times less sensitive than the nominal mode. 
}
\label{fig12}     
\end{figure}

\section{Summary}
\label{sec5}

SPIRE spectrometer bright-source mode calibration went through a major 
upgrade in March 2013 [i.e., starting in {\it Herschel} Interactive 
data Processing Environment (HIPE), version 11], by adopting a full 
nonlinearity correction scheme and integrating its data processing 
into the existing data processing pipeline of the nominal detector mode.  
This simplified data processing scheme requires only one additional 
phase-related gain correction procedure and two calibration products unique 
to the bright-source mode:  a nonlinearity correction product and a 
frequency-dependent, composite gain correction.  The derivations of 
these calibration products were given in this paper.  We have demonstrated 
that the bright-source mode flux calibration is within ${\sim}2\%$ of 
that of the nominal mode for their overlapping flux range, and agrees 
within a few percent with the model spectrum of Mars at flux levels as 
high as ${\sim}$10,000$\,$Jy, which is close to the targeted upper flux 
density calibration for the bright-source mode.
This represents a clear improvement over the 10\% flux consistency 
between the two detector modes achieved in the earlier calibration versions 
(i.e., HIPE 9 and 10).  We showed that 
the bright-source mode sensitivity is about 3-4 times less than that 
of the nominal mode.

\begin{acknowledgements}
We thank both an anonymous referee and Dr.~Locke Spencer for their useful comments
that helped improve the overall clarity of the paper. 
SPIRE has been developed by a consortium of institutes led by Cardiff University (UK) and including Univ. Lethbridge (Canada); NAOC (China); CEA, LAM (France); IFSI, Univ. Padua (Italy); IAC (Spain); Stockholm Observatory (Sweden); Imperial College London, RAL, UCL-MSSL, UKATC, Univ. Sussex (UK); and Caltech, JPL, NHSC, Univ. Colorado (USA). This development has been supported by national funding agencies: CSA (Canada); NAOC (China); CEA, CNES, CNRS (France); ASI (Italy); MCINN (Spain); SNSB (Sweden); STFC (UK); and NASA (USA).
The Herschel spacecraft was designed, built, tested, and launched under a contract to ESA managed by the Herschel/Planck Project team by an industrial consortium under the overall responsibility of the prime contractor Thales Alenia Space (Cannes), and including Astrium (Friedrichshafen) responsible for the payload module and for system testing at spacecraft level, Thales Alenia Space (Turin) responsible for the service module, and Astrium (Toulouse) responsible for the telescope, with in excess of a hundred subcontractors.
Support for this work was in part provided by NASA through an award issued by JPL/Caltech.

\end{acknowledgements}

%

\end{document}